\documentclass[RNAAS]{aastex62}

\begin{document}

\title{A Reader Friendly  Formalism for Circumstellar Material-Supernova Ejecta
	Interaction Model}

\correspondingauthor{Liang-Duan Liu; He Gao}
\email{liuliangduan@bnu.edu.cn, gaohe@bnu.edu.cn}

\author[0000-0002-8708-0597]{Liang-Duan Liu}
\affiliation{Department of Astronomy, Beijing Normal University, Beijing 100875, China}

\author{Ling-Jun Wang}
\affiliation{Astroparticle Physics, Institute of High Energy Physics, Chinese Academy of Sciences, Beijing 100049, China}

\author[0000-0002-3100-6558]{He Gao}
\affiliation{Department of Astronomy, Beijing Normal University, Beijing 100875, China}


\begin{abstract}
   The interaction of a SN ejecta with a pre-existing circumstellar material (CSM) is one of the most promising energy sources for a variety of optical transients. Recently, a semi-analytic method developed by Chatzopoulos et al. (2012, hereafter CWV12) has been commonly used to describe the optical light curve behaviors under such a scenario. We find that the expressions for many key results 
  in  CWV12 are too complicated for readers to make order of magnitude estimation or parameter dependency judgement.  Based on the same physical picture, here we independently re-derive all the formulae and re-establish a set of reader friendly formula expressions. Nevertheless, we point out and correct some minor errors or typos existing in CWV12. 
\end{abstract}
\section{Ejecta-CSM interaction model}

CWV12 combined the self-similar solutions presented by \cite{Chev1982} with \cite{A82} diffusion modeling formalism to calculate the bolometric optical light curves powered by CSM-ejecta interaction.

With the SN kinetic energy  $E_{\text{SN}}$ and the ejecta mass $M_{\text{ej}},$ the density profile could be expressed as \citep{Kasen2016}
\begin{equation}
\rho _{\text{ej}}\left( v,t\right) =\left\{
\begin{array}{lc}
\zeta _{\rho }\frac{M_{\text{ej}}}{v_{\text{tr}}^{3}t^{3}}\left( \frac{r}{v_{%
\text{tr}}t}\right) ^{-\delta }, & v<v_{\text{tr}}, \\
\zeta _{\rho }\frac{M_{\text{ej}}}{v_{\text{tr}}^{3}t^{3}}\left( \frac{r}{v_{%
\text{tr}}t}\right) ^{-n}, & v \geq v_{\text{tr}},%
\end{array}%
\right.
\end{equation}%
where the transition velocity $v_{\text{tr}}=\zeta _{v}\left( E_{\text{SN}}/ M_{\text{ej}}\right)
^{1/2} $ is obtained from the
density continuity condition \footnote{CWV12 took a scaling
parameter for the ejecta density profile as $g^n=[2(5-\delta)(n-5)E_{\mathrm{SN}}]^{(n-3)/2}/[(3-\delta)(n-3)M_{\mathrm{ej}}]^{(n-5)/2}/(4\pi(n-\delta))$. We find it can be simplified as $g^n=\zeta_{\rho}M_{\mathrm{ej}}v_{\mathrm{tr}}^{n-3}$.}. The outer density index $n$ depends on SN progenitor. For core-collapse SNe typical values are $\delta=1, n=10$ \citep{Kasen2016}. 
The coefficients $\zeta_\rho$ and $\zeta_v$ are given in Eqs (6) and (7) of \citep{Kasen2016}.  Since $v_{\text{tr}}$ cannot be observed directly, a constant $x_0$ is introduced to relate $v_{\text{tr}}$  and the characteristic velocity $v_{\text{SN}}$ of SN expansion, i.e. $v_{\text{SN}} = v_{\text{tr}}/x_0.$ The fraction of outer region ejecta mass is
\begin{equation}
    \xi_M=\frac{x_0^3-x_0^n}{\frac{n-3}{3-\delta}x_0^3+(x_0^3-x_0^n)}.
\end{equation}

Assume that the CSM density profile is:
\begin{equation}
\rho _{\text{CSM}}\left( r\right) =qr^{-s}.
\end{equation}
where $q$ is a scaling factor, which could be set as $q=\rho _{\text{CSM,in}%
}R_{\text{CSM,in}}^{s}$, where $\rho _{\text{CSM,in}}$ is the density at the CSM inner radius $R_{\text{CSM,in}}$. $s$ is the power-law index for the CSM density profile. $s=2$ is wind-like CSM profile and $s=0$ is shell-like CSM profile.

The forward shock(FS)/reverse shock(RS) dynamics could be described by a self-similar solution \citep{Chev1982}
\begin{eqnarray}
R_{\text{FS}} =R_{\text{CSM,in}}+\beta _{\text{FS}}t_{\text{tr}%
}v_{\mathrm{tr}}\left( \frac{t}{t_{\text{tr}}}\right) ^{\frac{n-3}{n-s} }  \label{Eq:RFS} \\
R_{\text{RS}} =R_{\text{CSM,in}}+\beta _{\text{RS}}t_{\text{tr}%
}v_{\mathrm{tr}}\left( \frac{t}{t_{\text{tr}}}\right) ^{\frac{n-3}{n-s}  } \label{Eq:RRS} ,
\end{eqnarray}%
here we introduce a characteristic time 
\begin{equation}
t_{\text{tr}}=\left( \frac{A\zeta _{\rho }M_{\text{ej}}}{q v_{\mathrm{tr}}^{3-s}}%
\right) ^{\frac{1}{3-s}},
\end{equation}
where $A$, $\beta _{\text{FS}}$, and $\beta _{\text{RS}}$ are constants that depend on the values of $n$ and $s$ which can be found in Table 1 of \cite{Chev1982}. Using $v_{\mathrm{FS}}=d R_{\text{FS}} /d t = v_{\mathrm{FS,tr}} (t/t_{\mathrm{tr}})^{(s-3)/(n-s)}$ yields the velocity of the FS.

The CSM mass swept up by the FS is
\begin{eqnarray}
M_{\text{FS,sw}}\left( t\right)  =\int_{R_{\text{CSM,in}}}^{R_{\text{FS}%
}\left( t\right) }4\pi r^{2}\rho _{\text{CSM}}\left( r\right) dr=M_{\mathrm{FS,tr}}\left( \frac{t}{t_{\text{tr}}}\right) ^{\frac{\left(
n-3\right) \left( 3-s\right) }{\left( n-s\right) }},
\end{eqnarray}
where $M_{\mathrm{FS,tr}}=4\pi A \zeta_\rho \beta_{\text{FS}}^{3-s}M_{\mathrm{ej}}/(3-s)$ is the FS swept up mass at $t_{\mathrm{tr}}$.

The FS input luminosity is
\begin{eqnarray}
L_{\text{FS}}\left( t\right) =\epsilon \frac{dE_{\text{K,FS}}}{dt} =\epsilon \frac{d}{dt}%
\left( \frac{1}{2}M_{\text{FS,sw}}{v}_{\text{FS}}^{2}\right)=L_{\mathrm{FS,tr}}\left( \frac{t}{t_{\text{tr}}}\right) ^{\gamma}.
\end{eqnarray}
where $\epsilon$ is an efficiency factor \footnote{CWV12 assumed $\epsilon=1$ that is  100\% efficiency in converting kinetic energy to radiation. It may be reasonable for $M_{\mathrm{CSM}} \sim M_{\mathrm{ej}}$, which is more likely for extreme luminosities in superluminous supernovae . But this assumption is unrealistic for $M_{\mathrm{CSM}} \ll M_{\mathrm{ej}}$ case. },  the temporal index $\gamma={(2n+6s-ns-15)} / {\left( n-s\right) } $ and the characteristic luminosity $L_{\mathrm{FS,tr}}\equiv\epsilon (\gamma+1)M_{\mathrm{FS,tr}}v_{\mathrm{FS,tr}}^2/ 2 t_{\text{tr}}. $ 

The FS energy input is  terminated when it breaks out to the optically thin CSM as
\begin{equation}
t_{\text{FS,BO}}=t_{\text{tr}} \left( \frac{ M_{\text{CSM,th}
}}{M_{\mathrm{FS,tr}}} \right) ^{ \frac{(n-s)}{ \left( n-3\right) \left( 3-s\right)} }.
\end{equation}
where $M_{\text{CSM,th}}$ is the optically  thick component of the CSM mass, the expression could be found in Eq(17) of CWV12.

Similarly, the ejecta mass swept up by RS is
\begin{equation}
    M_{\mathrm{RS,sw}}(t) = \int_{R_{\mathrm{RS}}(t)}^{R_{\mathrm{SN}}(t)} 4 \pi r^2 \rho_{\mathrm{ej}} d r=M_{\mathrm{RS,tr}}\left( \frac{t}{t_{\text{tr}}}\right) ^{\frac{\left(
n-3\right) \left( 3-s\right) }{\left( n-s\right) }},
\end{equation}
where $M_{\mathrm{RS,tr}}\equiv 4 \pi \zeta_\rho \beta_{\mathrm{RS}}^{3-n}M_{\mathrm{ej}}/(n-3)$. In the last term, we set $\rho_{\mathrm{ej}}\propto r^{-n}$ and ignore the contribution of inner shallow component, since the self-similar solution Eq (\ref{Eq:RRS})  is valid only for steep ejecta profile. The RS termination time $t_{\text{RS,*}}$ is determined when all the outer ejecta $\xi_M M_{\mathrm{ej}}$ has been swept, i.e,
\begin{equation}
t_{\text{RS,*}}= t_{\text{tr}}\left[ \frac{ \xi_M \left( n-3\right) }{4\pi \zeta _{\rho }\beta _{
\text{RS}}^{3-n}} \right] ^{\frac{n-s}{\left( 3-s\right)
\left( n-3\right) }}.
\end{equation}
In CWV12, they assumed that RS could swept all ejecta mass,  that is,  $\xi_M=1$. Actually,  for typical value of $x_0$ and $n$ one has $\xi_M \sim 0.2$, which means $t_{\text{RS,*}}$ was overestimated in CWV12.

In the comoving frame of the shock front of the homologously expanding
ejecta, the velocity of RS is \citep{Wang2019}
\begin{eqnarray}
\tilde{v}_{\text{RS}} &=&\frac{dR_{\text{RS}}}{dt}-\frac{R_{\text{RS}}}{t}=v_{\mathrm{RS,tr}}\left( \frac{t}{t_{\text{tr}}}\right) ^{\frac{s-3}{n-s}},
\end{eqnarray}
where $v_{\mathrm{RS,tr}}\equiv (s-3)\beta _{\text{RS}}v_{\mathrm{tr}}/(n-s)$.

The heating rate of the reverse shock is%
\begin{eqnarray}
L_{\text{RS}}(t) =\epsilon \frac{dE_{\text{K,RS}}}{dt}=\epsilon \frac{d}{dt}%
\left( \frac{1}{2}M_{\text{RS,sw}}\tilde{v}_{\text{RS}}^{2}\right)=L_{\mathrm{FS,tr}}\left( \frac{t}{t_{\text{tr}}}\right) ^{\gamma}
\end{eqnarray}
where the characteristic RS input luminosity $L_{\mathrm{RS,tr}}\equiv\epsilon (\gamma+1)M_{\mathrm{RS,tr}}v_{\mathrm{RS,tr}}^2/ 2 t_{\text{tr}}. $ 

We find that the FS/RS input luminosity follows the same behavior with time, then one has 
\begin{equation}
    \frac{L_{\mathrm{FS}}(t)}{L_{\mathrm{RS}}(t)}=A \left( \frac{n-3}{3-s} \right)^3 \beta_{\mathrm{RS}}^{n-5}\beta_{\mathrm{FS}}^{5-s}.
\end{equation}
The ratio of FS input luminosity to RS input luminosity only depends on the values of  $n$ and $s$. For $n=12$ and $s=0$, the value of ${L_{\mathrm{FS}}(t)} /{L_{\mathrm{RS}}(t)}=7.55$.   For $n=12$ and $s=2$, the value of ${L_{\mathrm{FS}}(t)} /{L_{\mathrm{RS}}(t)}=42.7$. This shows in wind case the contribution of RS is much smaller than that of FS at the early time.

Comparing the above equations with those in CWV12, we find the expressions for FS given by CWV12 are correct. The RS input luminosity in Eq (B7) of CWV12 misses a factor $(n-5)\beta_{\mathrm{RS}}^{5-n}/(n-3)$ \footnote{For sensible choices of the density profile slope $n$ and $\beta_{\mathrm{RS}}$, this factor is about one, so the change is unlikely to alter the published results within the involved uncertainties and parameter degeneracy.}.  It is worth noting that the self-similar power-law solutions Eqs (\ref{Eq:RFS}) and (\ref{Eq:RRS}) only holds for times $t \leq t_{\mathrm{tr}}$ when the interacting regime  remains in the outer ejecta.

\acknowledgments
LD is supported by China Postdoctoral Science Foundation (grant No. 2019M660515 and BX20190044).

\end{document}